\documentclass[10pt,twocolumn]{article}

\usepackage[a4paper,margin=1.9cm]{geometry}

\usepackage{xurl}              
\usepackage{hyperref}         
\hypersetup{hidelinks}        

\usepackage{lmodern}        
\usepackage[T1]{fontenc}
\usepackage{microtype}
\usepackage{placeins}
\usepackage{algorithm}
\usepackage{algpseudocode}
\usepackage{booktabs}
\usepackage{threeparttable}  

\usepackage{amsmath,amssymb,amsthm}
\usepackage{graphicx}

\usepackage{booktabs}      

\usepackage{titlesec}

\titleformat{\section}
  {\normalfont\bfseries\large}
  {\thesection.}
  {0.5em}
  {}

\titleformat{\subsection}
  {\normalfont\bfseries\small}
  {\thesubsection.}
  {0.5em}
  {}

\titleformat{\subsubsection}
  {\normalfont\itshape\small}
  {\thesubsubsection.}
  {0.5em}
  {}

\usepackage[numbers,sort&compress]{natbib}

\usepackage{subfiles}
\usepackage{amsthm}

\theoremstyle{plain}

\theoremstyle{definition}

\theoremstyle{remark}

\usepackage{graphicx}

\usepackage[font=footnotesize,labelfont=bf]{caption}
\usepackage{subcaption}   

\usepackage{authblk}       

\graphicspath{{Fig/}}

\title{Manipulation in Prediction Markets: An Agent-based Modeling Experiment}
\author[1,2,*]{Bridget Smart}
\author[1,*]{Ebba Mark}
\author[3]{Anne Bastian}
\author[4]{Josefina Waugh}

\affil[1]{\textit{Institute for New Economic Thinking, University of Oxford, Manor Road, OX1 3UQ, Oxford, United Kingdom}}
\affil[2]{\textit{Mathematical Institute, University of Oxford, Woodstock Road, OX2 6GG, Oxford, United Kingdom}}
\affil[3]{\textit{Institute for Globally Distributed Open Research and Education (IGDORE), Gothenburg, Sweden}}
\affil[4]{\textit{Economics Institute, Pontifical Catholic University of Chile, Avenida Libertador Bernardo O'Higgins 340, Santiago, Chile}}

\date{} 

\begin{document}

\twocolumn[
  \begin{@twocolumnfalse}
    \maketitle
    \begin{abstract}
      Prediction markets mobilize financial incentives to forecast binary event outcomes through the aggregation of dispersed beliefs and heterogeneous information. Their growing popularity and demonstrated predictive accuracy in political elections have raised speculation and concern regarding their susceptibility to manipulation and the potential consequences for democratic processes. Using agent-based simulations combined with an analytic characterization of price dynamics, we study how high-budget agents can introduce price distortions in prediction markets. We explore the persistence and stability of these distortions in the presence of herding or stubborn agents, and analyze how agent expertise affects market-price variance. Firstly we propose an agent-based model of a prediction market in which bettors with heterogeneous expertise, noisy private information, variable learning rates and budgets observe the evolution of public opinion on a binary election outcome to inform their betting strategies in the market. The model exhibits stability across a broad parameter space, with complex agent behaviors and price interactions producing self-regulatory price discovery. Second, using this simulation framework, we investigate the conditions under which a highly resourced minority, or ``whale'' agent, with a biased valuation can distort the market price, and for how long. We find that biased whales can temporarily shift prices, with the magnitude and duration of distortion increasing when non-whale bettors exhibit herding behavior and slow learning. Our theoretical analysis corroborates these results, showing that whales can shift prices proportionally to their share of market capital, with distortion duration depending on non-whale learning rates and herding intensity.
    \end{abstract}

    \vspace{1ex}


    \vspace{4ex}
  \end{@twocolumnfalse}
]

\begingroup
  \renewcommand{\thefootnote}{\fnsymbol{footnote}} 
  \footnotetext[1]{Corresponding authors:
    Bridget Smart, \texttt{bridget.smart@maths.ox.ac.uk}, and
    Ebba Mark, \texttt{ebba.mark@magd.ox.ac.uk}.}
\endgroup

\renewcommand{\thefootnote}{\arabic{footnote}}

\section{Introduction}

Prediction markets, where participants wager on binary outcomes, have been proposed as effective mechanisms for aggregating dispersed beliefs into accurate predictors in environments characterized by uncertainty and heterogeneous access to information \citep{sethiPoliticalPredictionWisdom2025,allenBeautyContestsIterated2006}. Their appeal rests on the idea that traders, motivated by financial incentives, reveal private information through their willingness to buy or sell contracts corresponding to a particular outcome. 

Prediction market prices have been found to closely approximate the mean belief about an event’s probability in theoretical settings with well-informed, constrained and risk-adverse traders \citep{wolfersInterpretingPredictionMarket,gjerstadRiskAversionBeliefs2004}. In several real elections, prediction-market forecasts have matched or outperformed traditional polling or expert judgment \citep{rothschildForecastingElections2009,bergPredictionMarketAccuracy2008, wingard_polymarket_2025, cutting_are_2025}, although evidence is mixed \citep{leighCompetingApproachesForecasting2006, erikson_markets_2012}. Positive examples are sometimes attributed to the ``wisdom of crowds'', which relies on assumptions of independence, diversity, and the absence of dominant participants, conditions which are not always satisfied in practice \citep{surowieckiWisdomCrowdsWhy2004}.

Much of the empirical evidence supporting prediction-market accuracy comes from markets with strict limits on individual trading positions. Early studies of prediction market accuracy and dynamics largely relied on platforms with strict per-trader position limits, such as the Iowa Electronic Markets \citep{bergPredictionMarketAccuracy2008} and PredictIt \citep{restocchiOpinionDynamicsExplain2023}, where individual exposure was constrained to a few thousand U.S. dollars per market (e.g., \$3,500 on PredictIt as of 2025 \citep{thomson_reuters_practical_law_cftc_2025_2}). These caps strictly limited any single participant’s ability to meaningfully influence prices. 

In contrast, newer platforms permit substantially larger, contract-specific position limits for eligible participants, in some cases reaching the multi-million-dollar range. Under these conditions, the classic assumption that individual traders are too small to affect prices, and that any attempted manipulation is rapidly arbitraged away \citep{hansonCombinatorialInformationMarket}, is no longer guaranteed to hold. Both theoretical and empirical work shows that when traders are sufficiently capitalized or face strong incentives, prediction markets may sustain non-informational price deviations for meaningful periods \citep{camererCanAssetMarkets1998,allenBeautyContestsIterated2006}.

This institutional shift has coincided with a rapid expansion in prediction market activity. During the 2024 U.S. presidential election, market capitalization reached hundreds of millions of USD \cite{kalshiData2025,ExploreHighestVolume}. For example, weeks prior to the 2024 U.S. presidential election between Donald Trump and Kamala Harris, a consequential ruling by the Commodity Futures Trading Commission legalized KalshiEx \citep{united_states_court_of_appeals_for_the_district_of_columbia_circuit_kalshiex_2024}. Within days, the platform announced its ability to accommodate individual trades of up to \$100 million \citep{garrettAmericansBet1002024, mansour_its_2024}, with notional volume exceeding \$20 billion in December 2025 \cite{kalshiData2025}, marking a departure from smaller, tightly-capped markets. 

In this context, a type of market distorter commonly referred to as a ``whale''\footnote{``Whales'' refer to participants in a betting market who control and deploy a volume of capital significantly larger than the average participant, thereby potentially possessing the ability to temporarily influence market prices through their trades or wagers.} has been identified across multiple platforms in recent years \citep{quiroz-gutierrez_french_2024, foxWhatKnowPotential2024}.  For example, in 2024, Polymarket identified a \$45 million bet placed by a French national to favor Donald Trump, temporarily pushing up his odds in the prediction market \citep{quiroz-gutierrez_french_2024, sorkin_french_2024}. At such levels of capital concentration, salient price movements may interact with heterogeneity in how traders weight prior beliefs, market-external information, and market prices, generating dynamics that differ qualitatively from those observed in smaller, tightly capped markets.

The potential for such market manipulation invites a suite of questions regarding the potential feedback between prediction markets and real-world political outcomes. Their credibility as predictive tools could earn them a higher rank in the information landscape available to voters when shaping their voting decisions. Research on informational cascades and bandwagon effects suggests that shifts in public beliefs triggered by prominent information signals like polls can influence real political outcomes, making endogeneity between market prices and behavior difficult to rule out \citep{mortonWhatMotivatesBandwagon2015, morwitz_polls_1996, boukouras_can_2023, alabrese_national_2022, rothschild_are_2014}. Anecdotal and journalistic evidence indicates that members of the public have begun to rely on prediction market values in addition to polls to form their understanding of public opinion \citep{escande_us_2024, garrettAmericansBet1002024}. Thus, a potential feedback between market price and electorate beliefs raises a logical concern about incentives for strategic manipulation of market price to potentially distort democratic processes.

This motivates our central question: \textbf{Under what conditions do contemporary prediction markets function as self-correcting information processing mechanisms, and when can they instead temporarily sustain price deviations introduced by a highly resourced, biased minority?} 

We address this by examining the prediction market as an information-aggregating mechanism and how trader behavior can introduce volatility and error into the market. In particular, we consider how large trades can lead to sustained or fluctuating price movements through a biased large trader (whale) operating among traders who partially herd toward the market price.

\subsection{Methodological Approach and Contributions}
In this study, we develop an open source Agent-Based Model (ABM) which simulates a prediction market in which betting agents, characterized by heterogeneous learning functions, expertise, biased perceptions, and budgets, bet on the outcome of a binary election. We use this model to consider how wealth distribution, risk aversion, herding and belief dynamics influence prediction market price. Our proposed agent-based model framework exhibits stable performance and accurately predicts the data-informed stylized election outcome across a wide range of parameter values. Furthermore, a decision rule wherein bettors aim to maximize risk-adjusted returns produces reasonable outcomes in terms of prediction market price accuracy (in relation to leading election forecasting models) and where profits accrue to those bettors with more accurate insight into the electoral outcome.

Next, using this model we investigate the relative profit gain between experts and non-experts, the potential for market price manipulation by high-capital investors, and whether the stability of the market depends on the incorporation of herding behavior. We evaluate the performance of this prediction market, defined as the deviation of the market price from public voting preferences, using a regression model to test for lagged correlation dynamics. We demonstrate that, over a particular population size, initial capital allocation, and wide range of parameter values, the market quickly adjusts to counteract the price pressure of whale bettors. However, whale bettors are able to temporarily shift and introduce volatility into the market price, where the magnitude and duration of this shift is proportional to the product of their budget allocation and misvaluation. The introduced error can be amplified in the presence of herding or stubborn behavior from other bettors.

Alongside insights from our ABM, we conduct a theoretical analysis of agent valuation, utility, and market-update functions to examine how the prediction market responds to the presence of whales and herding agents. This analysis reinforces our simulation results, showing that prolonged whale-induced price distortions scale with the whale’s share of total market capital. The decay of these distortions depends on the learning dynamics and herding propensity of other agents. We also demonstrate that market distortions may decay slowly or oscillate in a market with slow belief updates or herding toward the market price.

Beyond this investigation of prediction markets as information aggregators, we contribute analytical tools for future research by making our ABM model open source, available on \href{https://github.com/ebbam/power_prediction/}{GitHub}\footnote{\url{https://github.com/ebbam/power_prediction/}}. Additionally, we build a Dash application providing a graphical user interface to the ABM, allowing users to change parameters for agent behavior and investigate the impact of correlations within the model through a graphical user interface. The modular ABM provides considerable flexibility for extensions that could accommodate alternative behavioral rules, parameter variation, exploration of the role of information shocks, and more detailed models of an election process in which the network of an electorate impacts the evolution of voting preferences.

Overall, this work has implications for the regulation and design of prediction markets at a time when their legal contexts are evolving. Our findings motivate consideration of measures that limit potential impacts of concentrated capital, particularly in election prediction market settings where prices may function as public information signals, including restrictions on individual trade sizes or overall budget limits where enforceable. While feedback effects between prediction markets and democratic processes remain empirically unresolved, our results add weight to a precautionary approach.  Beyond market design, our findings highlight risks associated with the public framing of election betting market prices as probability forecasts, especially when such prices may be influenced by concentrated capital and subsequently amplified by media coverage or interpreted by voters, campaigns, or other downstream audiences.

\subsection{Related Literature}

Existing research on prediction markets spans theoretical, experimental, and simulation-based approaches. A substantial body of work examines prediction markets as forecasting tools, drawing on their modern institutional advantages \cite{yehUsingPredictionMarkets2006,polgreenUsePredictionMarkets2007,sethiPoliticalPredictionWisdom2025,rothschildForecastingElections2009,spannSportsForecastingComparison2009}, theoretical motivations such as the efficient market hypothesis \cite{williamsPredictionMarketsTheory2019,bergPredictionMarketAccuracy2008}, empirical evidence of forecasting accuracy \cite{forsytheAnatomyExperimentalPolitical1992,bergPredictionMarketAccuracy2008}, and social-sensing perspectives \cite{galesicHumanSocialSensing2021}, even as some foundational assumptions have been challenged \cite{danaAreMarketsMore2019,leighCompetingApproachesForecasting2006}. Experimental studies have probed these mechanisms in controlled settings, examining how institutional design and participant incentives shape price formation \cite{vanbruggenPredictionMarketsInstitutional2010,qiuTwitterBasedPredictionMarket2011}. Notably, \cite{hansonInformationAggregationManipulation2006} showed that small-scale markets can be resistant to manipulation, while \cite{boulu-reshefRiskAversionPrediction2016} linked price dynamics to participants’ risk aversion.

A growing literature employs agent-based or computational models to explore prediction-market dynamics beyond what analytical or experimental methods permit. \cite{klingertComparingPredictionMarket2012} Use a multi-agent simulation to evaluate alternative market mechanisms, and \cite{rothschildTradingStrategiesMarket2013} couple empirical data with a single stylized trader to study how individual trading strategies influence performance. Other work embeds prediction-market behavior within social or informational networks, examining how opinion dynamics \cite{restocchiOpinionDynamicsExplain2023}, social-media-derived signals \cite{bothosUsingSocialMedia2010}, or simulated electorates \cite{yuAgentBasedModelingPrediction2012} shape prices. \cite{shenRoleWhaleInvestors2025} develop an agent-based model to study how herding behavior and social network structure interact with agent heterogeneity to generate price volatility in Bitcoin markets.

Unlike prior work, our simulation-based and analytic approach models a prediction market with a well-defined true price and an exogenous noisy information signal. We explicitly incorporate heterogeneous expertise, behavioral learning rates, and budget dispersion, enabling systematic analysis of how behavioral and financial asymmetries affect price stability. By decoupling social-network effects, we can directly characterize how behavioral heterogeneity, noisy information, and concentrated wealth interact to induce persistent distortions and affect market stability.

\section{Agent-based Model}\label{sec:model}

We introduce an open-source agent-based model (ABM) simulating a prediction market in which $N$ agents, indexed $i=0,1,\dots N$, trade contracts placing bets on the outcome of a binary election over $T$ discrete time periods indexed $t=0,1,\dots T$. The model for the betting agent's behavior is inspired by \cite{sethiEvaluatingPredictionMechanisms2024}. In this section, we outline the market structure, agents, and actions taken on the market. Simulation \ref{sim:abm_pred_market} outlines the full simulation process in pseudo-code. 

\floatname{algorithm}{Simulation}

\begin{algorithm}
  \caption{Simulation procedure for the agent-based price-first prediction market}
  \label{sim:abm_pred_market}
  \begin{algorithmic}[1]
    \State \textbf{Input:} betting agents $\mathcal{A}$, election time $T$, initial market price $m_0$, outcome uncertainty $\sigma^2_{\eta}$
    \State \textbf{Output:} market price trajectory $m$, agent budgets $B$, portfolios $C$, valuations $V$, and profit histories
    \State Initialize $t \leftarrow 0$
    \State Initialize  $\eta_0 = m_0$
    \While{$t < T$}
      \State $\theta_t \leftarrow []$ \Comment{Generate empty order book}
      \For{each betting agent $a \in \mathcal{A}$}
        \State Receive private signal $M_{a,t}$ \Comment{Agent receives fuzzy observation of outcome (\autoref{eq:signal_draw})}
        \State Propose order volume $x_{a,t}$ according to \autoref{eq:EU_fun_general} 
        \If{$x_{a,t} \neq 0$}
          \State Append order $x_{a,t}$ to order book $\theta_t$  
        \EndIf
      \EndFor
      \State Arrange order book in random order 
      \State Select matched orders, partially filling orders where necessary
      \State Execute matched orders $\theta^\star$ \Comment{Execute matched trades by transferring contracts and cash}
      \State Calculate net demand $D_t$ normalized by total order volume
      \State Update market price $m_{t} \leftarrow m_{t+1}$ as in \autoref{eq:price_update}
      \For{each agent $a \in \mathcal{A}$}
        \State Update wealth and holdings $(B_{a,t+1}, C_{a,t+1})$
        \State Update market valuations $V_{a,t+1}$ \Comment{\autoref{eq:valuation_update}}
      \EndFor
      \State $t \leftarrow t + 1$
      \State $\eta_{t} \leftarrow \eta_{t+1}$ \Comment{Update true election outcome \autoref{eq:election_outcome}}
    \EndWhile
  \end{algorithmic}
\end{algorithm}
\vspace{-3mm}
\subsection{Election}

At each time step $t$, agents bet on the outcome of a binary election outcome $\eta_t$. We represent the true election outcome $\eta_t$ as a random walk:
\begin{equation}\label{eq:election_outcome}
\eta_t = \eta_{t-1} + \epsilon_\eta , \quad \epsilon_{\eta} \sim \mathcal{N}(0, \sigma^2_{\eta})
\end{equation}
where $\eta_t$ is bounded by [0,1]. $\eta_t$ is a reflection of public opinion at time $t$ with no knowledge of future movements. 

\subsection{Market}

At each time step, betting agents submit buy or sell orders of a fixed size based at the current market price $m_t$. A central market fulfills these buy and sell orders via random order matching. The market has a bid-ask spread of zero, with a single price acting as both the buy and sell price. Orders are matched using double auction rules. Bettors update their budget and portfolio given orders filled. 

The market is responsive to net demand in the order book, setting the next time period's market price $m_{t+1}$ according to \autoref{eq:price_update}: 
\begin{equation}\label{eq:price_update}
m_{t+1} \;=\;  m_t \;+\; \lambda\; \dfrac{D_t}{K_t},
\end{equation}

where $D_t$ denotes the net demand, $K_t$ the total order volume, and scaling parameter $\lambda$, controlling the maximum step size of a single-$t$ price update. The market price is constrained to lie within $[0,1]$ as in real-world prediction markets.

\begin{table*}[ht]
\caption{Inventory of Bettor Characteristics and Behaviors\label{tbl:bettor_attributes}}
\tabcolsep=0pt
\begin{tabular*}{\textwidth}{@{\extracolsep{\fill}}lc p{7cm} c@{\extracolsep{\fill}}}
\toprule
\hline
\textbf{Characteristics} & \textbf{Symbol} & \textbf{Definition} & \textbf{Initial Value} \\[2pt]
\hline
\textbf{Resources} & & & \\
Budget & $B_{i,t}$ & Betting agent's budget at time $t$ & $B_{i,0} \sim U(100,1000)$\\
Market valuation & $V_{i,t}$ & Bettor's belief about the market value of a particular contract between \$0-1. & $V_{i,0} \sim N(0.5,0.05)$\\
Portfolio & $C_{i,t}$ & Holding position of each betting agent. A positive (negative) value of $C_{i,t}$ indicates that the betting agent has purchased more (less) contracts than they have sold. & $C_{i,0} = 0$ \\
\hline
\textbf{Attributes} & & & \\
Stubbornness & $s_i$ & Bettor's resistance to updating their internal market valuation $V_{i,t}$ when provided with new information. & $s_i \sim N(0.3,0.05)$ \\ 
Expertise & $e_i$ & This parameter controls how ``clearly'' the betting agent can see the true election outcome at each time step. & $e_i \sim N(0.9,0.04)$  \\  
Bias & $b_i$ & This parameter reflects the extent to which the bettor systematically over- or under-estimates the true election outcome. Alternatively, this bias reflects an internal belief that diverges from the true market value signal the bettor receives at each time step. & $b_i \sim 0$ \\
Risk aversion & $r_i$ & Bettor's risk preferences. & $r_i \sim U(0,1)$\\
\bottomrule
\end{tabular*}
\begin{tablenotes}%
 \footnotesize{\item Note: Attributes are all constrained to be between 0 and 1, so clipping is performed to ensure values are valid.
\item $^{1}$ Some of the default values for these parameters are drawn from a uniform ($U$) or normal ($N$) distribution, while others are constant.
\item $^{2}$ These initial values were chosen to ensure heterogeneous yet realistic agent behavior, providing sufficient diversity in beliefs and trading strategies while maintaining market dynamics that remain consistent with the true election outcome.}\vspace*{6pt}
\end{tablenotes}
\end{table*}

\subsection{Agents}

The betting agents are defined by the following characteristics and behaviors. 

\subsubsection{Characteristics}

First, each betting agent possesses the characteristics listed in \autoref{tbl:bettor_attributes}, jointly determining how they behave on the prediction market. Attributes are fixed and specified uniquely for each agent at model initialization. Resources change throughout the course of the simulation following market activity.

\subsubsection{Behavioral Rules}

\paragraph{Internal Valuation}

At each time step $t$, each betting agent updates their individual valuation $V_{i,t}$ of the true election outcome $\eta_t$. This can analogously be considered the betting agent's individual belief about $\eta_t$. 

The agent does not observe $\eta_t$ directly. Rather, they receive a ``fuzzy'' signal $M_{i,t}$ of the true election outcome $\eta_t$ whose variance decreases with their expertise $e_i$. Agents with an expertise value closer to 1 will have an expected individual valuation with less variance around the market price, and as $e_i \to 0$, the market signal becomes increasingly noisy. This is implemented using a normal distribution centered at $\eta_t$ with variance $1-e_i$. Additionally, each betting agent has a bias $b_i$ which they detract from signal $M_{i,t}$ when updating their internal valuation. In this set-up, the bias impacts the update value consistently at each time step. 

Bettors update their internal valuation of the market price as outlined in \autoref{eq:valuation_update}.
\begin{equation}\label{eq:valuation_update}
V_{t+1} = (1-s_i) (M_{i,t} - b_i) + s_iV_{i,t}
\end{equation}
\begin{figure}[H]
    \centering
    \includegraphics[width=\linewidth]{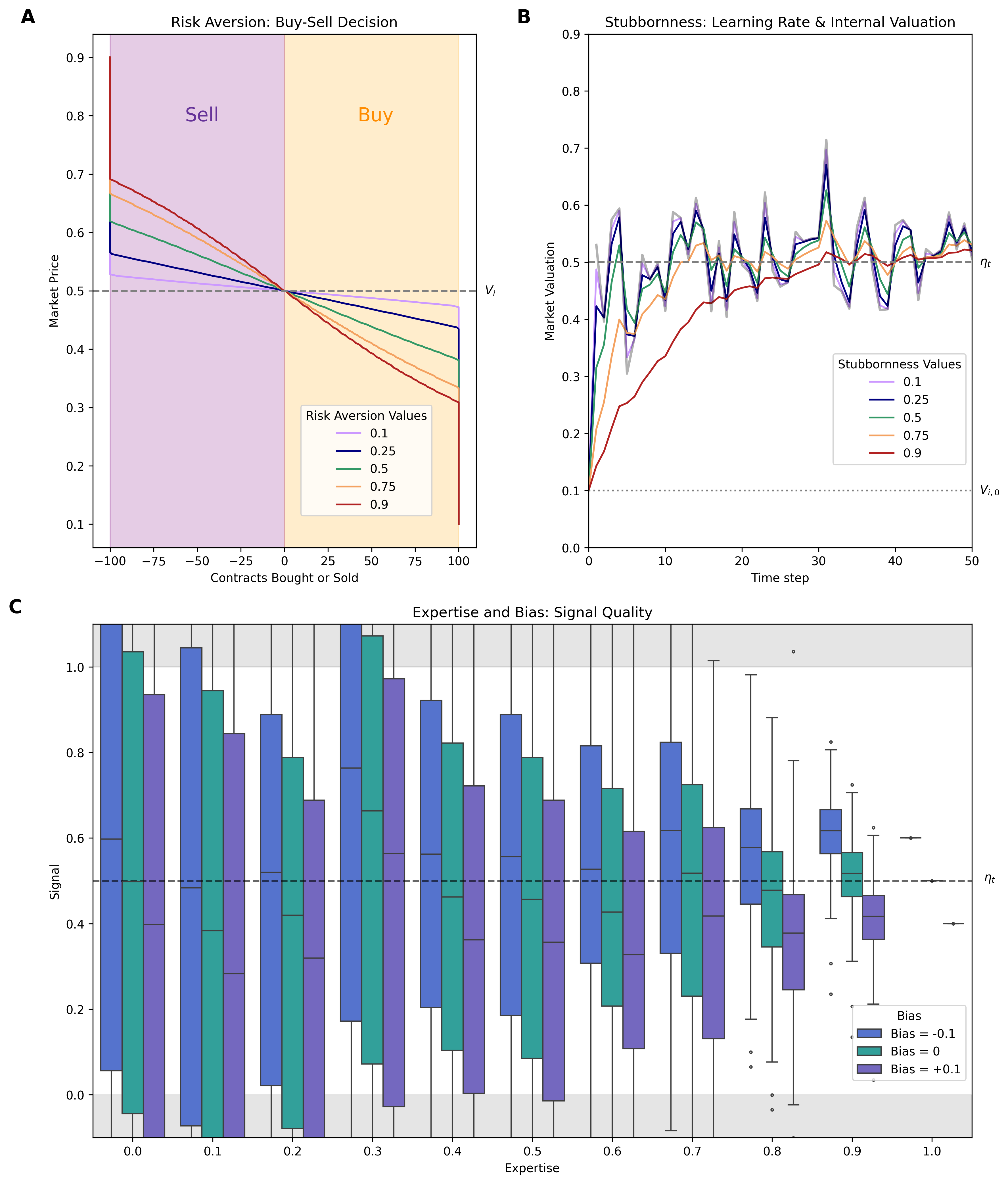}
    \caption{{Stylized representation of behavioral attributes and their effects on betting agent behavior. \textbf{Panel (A)} represents the buy-sell decision of an individual agent $i$ at time $t$ with budget $B_{i,t} = \$100$ whose internal market valuation $V_{i,t} = \$0.50$. Each line represents their buy-sell decision at various market prices $m_t$, varying their degree of risk aversion where $r_i \rightarrow 0$ implies higher levels of risk aversion. \textbf{Panel (B)} represents a high-expertise ($e_{i} = 0.9$) betting agent's learning process over 50 time steps. The agent starts the simulation with an internal valuation $V_{i,0} = \$0.10$ receiving a signal of the true value $\eta_t$ (a constant value in this stylized representation). Each line represents the learning trajectory mediated by varying stubbornness $s_i$. \textbf{Panel (C)} represents the distribution of signal $M_{i,t}$ where the true value $\eta_t = \$0.50$, varying expertise $e_i$ and bias $b_i$. Each color represents a different level of bias $b_i$ and each value on the x-axis represents a different level of expertise $e_i$. The gray sections represent the limits ([\$0,\$1]) of any market valuation.}}
    \label{fig:stylised_behaviours}
\end{figure}
where
\begin{equation}\label{eq:signal_draw}
M_{i,t} \sim \mathcal{N} (\eta_t, \;1-e_i)
\end{equation}

and $V_{t-1}$ is constrained between [0,1]. Conditional on $V_{i,t}$ and $\eta_t$,
\begin{align}
V_{i,t+1}\mid V_{i,t},\eta_t \sim \mathcal{N}\!\big(&(1-s_i)(\eta_t-b_i)+s_iV_{i,t},\nonumber\\&\;(1-s_i)^2(1-e_i)\big).
\end{align}

$V_{i,t+1}$ is then set to the boundary value if it falls outside of the interval $[0,1]$.

Whereas bias $b_i$ and expertise $e_i$ affect the quality and interpretation of signal $M_{i,t}$, respectively, stubbornness $s_i$ affects the agent's learning rate (\autoref{fig:stylised_behaviours} (B) and (C)). Agents are heterogeneous in their willingness to update their market valuation based on this signal of the true election outcome. We represent this stubbornness $s_i$ as a time-invariant bettor characteristic that affects their relative valuation of signal $M_{i,t}$ and their previous valuation $V_{i,t}$. Agents whose $s_i$ are close to $0$ mainly form their new market valuation in response to the most recent information signal, whereas agents whose $s_i$ are close to $1$ only incrementally change their valuation in light of new information.

This formulation aligns with canonical adaptive expectations or Bayesian belief updating models that take into account ``sticky'' or persistent beliefs in opinion or belief formation \citep{ortoleva_alternatives_2024, barron_belief_2021, burton_optimism_2022, aydogan_how_2025, henckel_belief_2022}.

\paragraph{Utility Function \& Order Placement}
Next, given their internal valuation $V_{i,t}$, budget $B_{i,t}$, portfolio holdings $C_{i,t}$, and market price $m_t$, each agent then proposes an optimal order volume $x^\star_t$. Agent $i$'s proposed order volume at time $t$ is given by the value $x^\star_t$ which maximizes the agent's expected utility function subject to their budget constraint. 
\begin{equation}\label{eq:EU_fun_general}
x^\star_t = \arg max_{x_t} E[u].
\end{equation}
Where
\begin{align}
E[u] = &V_t \;u ( B_t - m_t x_t + C_t + x_t )\nonumber \\&+ (1- V_t) \;u(B_t - m_t x_t),\label{eq:EU_fun}
\end{align}
subject to
\begin{equation*}
    0 \leq B_t \qquad \text{ and }\qquad
    0 \leq B_t + C_t,
\end{equation*}
where $u(\cdot)$ is a utility function defined by constant relative risk aversion for agents with heterogeneous risk preferences as in \autoref{eq:crra_utility_fun} \citep{phelps_users_2024}. A visualization is shown in \autoref{fig:stylised_behaviours} (A).
\begin{equation}\label{eq:crra_utility_fun}
u(\omega)=
\begin{cases}
\dfrac{1}{1-r_i} \omega^{1-r_i},\;\; \text{ if } r_i \geq 0, r_i \neq 1\\
\log(\omega), \;\; \text{ if } r_i = 1.
\end{cases}
\end{equation}

\paragraph{Herding Behavior}

So far we do not allow the market price to impact the internal valuation of agents. A substantial literature studies herding and social-learning mechanisms in financial markets. However, empirically distinguishing herding from correlated trading driven by shared external information remains difficult, and definitions of herding vary across studies \cite{dohertyInformationalCascadesFinancial2018, bikhchandani_herd_2001}. Rather than attempting empirical identification, we model herding as a behavioral response to the observed absolute market price. This choice is motivated by the design of binary prediction markets, where prices are commonly interpreted as salient probability signals, in contrast to the relative-price dynamics emphasized in equity markets. We therefore introduce an alternative formulation of \autoref{eq:valuation_update} incorporating an additional signal of market value in \autoref{eq:valuation_update_herding}. $h_i$ is the agent level parameter which controls the strength of herding, where $h_i\in[0,1]$.
\begin{equation}\label{eq:valuation_update_herding}
V_{i,t+1} = (1-h_i) \big(\big(1-s_i\big) \big(M_{i,t} - b_i\big) + s_iV_{i,t} \big)+ h_i \big(m_t)
\end{equation}

In \autoref{eq:valuation_update_herding}, individuals weight between the valuation calculated as in \autoref{eq:signal_draw} with the market price by $h_i$. We assume independent effects between $h_i$ and $s_i$.

\begin{figure*}
    \centering
    \includegraphics[width=\textwidth]{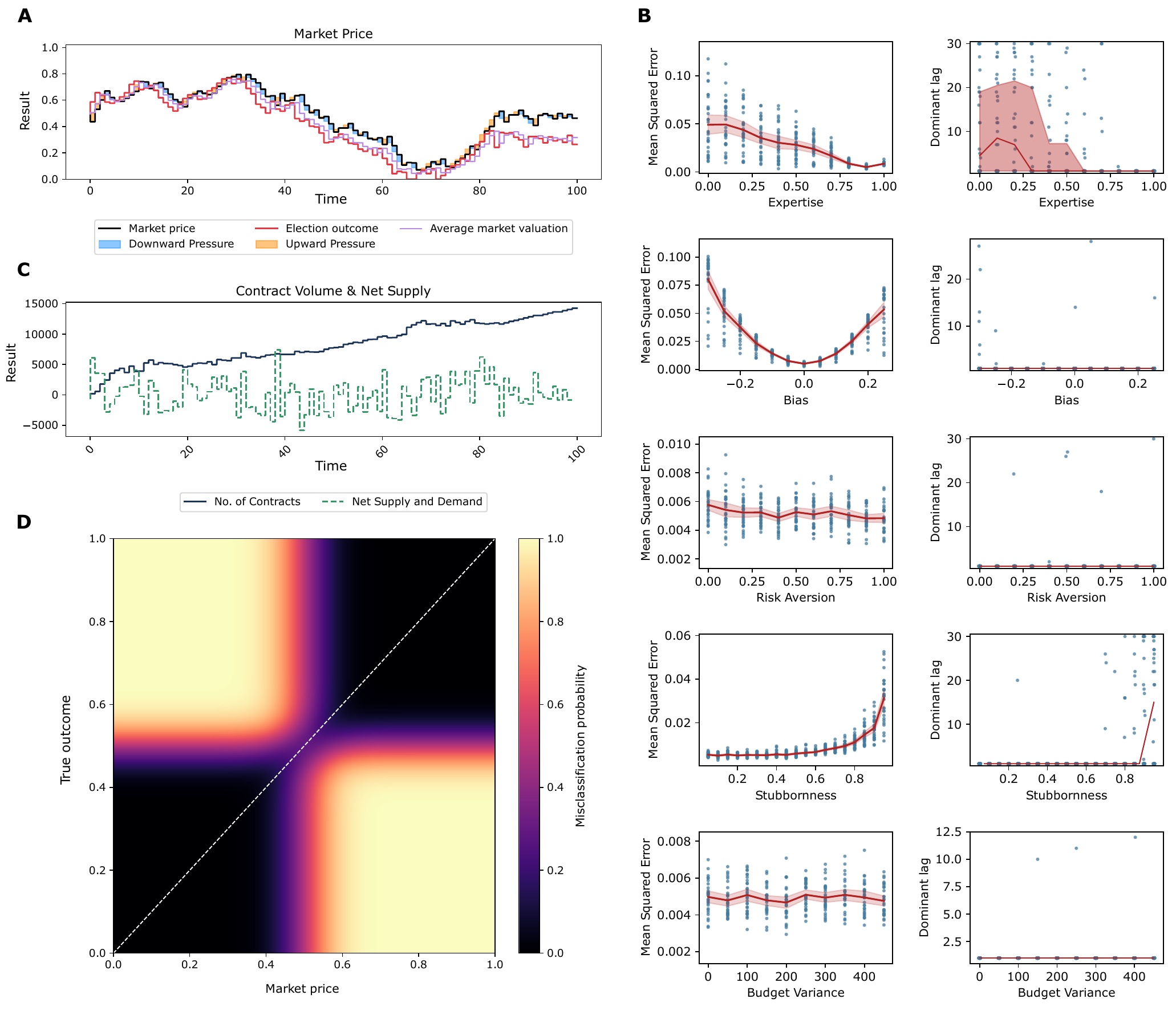}
    \caption{Panel A and C show a single simulation from the ABM. Panel A has the true election outcome (red), market price (black) alongside the net supply and demand indicating upward or downward pressure on the market price. The average market valuation of bettors is shown in purple. Panel C visualizes this single run through the number of contracts held across all agents (navy), as well as the net supply and demand (green). Panel (B) shows the validation and robustness checking across the bettor characteristics of expertise, bias, risk aversion, stubbornness and the variance of budgets. The left column shows the mean squared error introduced between the true outcome and the market price across a range of parameter values, and the right column shows the dominant lag. The dominant lag is the single lag $\ell$ whose one-variable regression of the outcome on the lagged market series yields the strongest statistical fit (lowest p-value). Each simulation has 100 agents over 100 time steps, with an initial price of 0.5 and the variance of the true election outcome 0.05. For each parameter value 30 simulations were performed and default values are $B_{i,0}\sim U(100,1000)$, $V_{i,0}\sim N(0.5,0.05)$, $C_{i,0}=0$, $s_i\sim \mathcal{N}(0.3,0.05)$, $e_i \sim \mathcal{N}(0.9,0.04)$, $r_i \sim U(0,1)$ as shown in \autoref{tbl:bettor_attributes}. Red lines show the average value across the simulations with a 95\% empirical confidence interval. Panel D shows the misclassification probability across different market price and true outcome pairs. As these values are further apart or closer to 0.5, the misclassification probability increases. The diagonal line marks perfect agreement between market and outcome. Regions near $\eta_t = 0.5$ and $m_t=0.5$ show a smooth transition between correct and incorrect classification. This transition becomes wider when uncertainty (noise) in either the market price or the true outcome increases, reflecting reduced confidence in whether the market’s implied prediction aligns with the eventual result.}
    \label{fig:F1}
\end{figure*}
\subsection{Modeling Assumptions}\label{sec:model_assumptions}

This model set-up is defined by the following model assumptions, further influencing our theoretical result outlined in \autoref{sec:theoretical_derivation} later in this work.

\begin{enumerate}
    \item \textbf{Binary contract payoff:} Each contract pays \$1 if the target event occurs and \$0 otherwise.
    \item \textbf{Agent information and signals:} Agents are indexed by $i=1,\dots,N$. At each timestep $t$ agent $i$ receives a private signal
    \[
    M_{i,t}\sim\mathcal{N}(\eta_t,\;1-e_i),
    \]
    where $\eta_t$ is the underlying (``true'') election value. Signals are assumed independent across agents conditional on $\eta_t$.
    \item \textbf{Price-taking agents (no self-impact):} Agents do not account for their own price impact when choosing order sizes. In other words, each agent optimizes while treating the market price $m_t$ as exogenous.
    \item \textbf{Order matching:} The ABM implements randomized order matching. For the analytic derivations below we assume all submitted orders are matched at the market price $m_t$. Under both cases the agent's expected payoff is consistent with a guaranteed execution at $m_t$.
    \item \textbf{Clipping / boundedness:} All prices $m_t$ and individual valuations $V_{i,t}$ are constrained to the open interval $(0,1)$. Any intermediate values outside of $(0,1)$ are clipped back into $(0,1)$ (hard clipping).
    \item \textbf{Feasibility / wealth domain:} Agents maintain cash $B_{i,t}\ge0$ and holdings $C_{i,t}\in\mathbb{R}$. The utility function requires post-trade cash/wealth arguments to be strictly positive.
    \end{enumerate}

\subsection{Validation and robustness}

Next, we provide evidence of the model's stability and accuracy across the parameter space by varying the distribution of bettor attributes before applying the ABM to investigate price distortions. Accuracy in this and subsequent sections is defined as the distance between the market price $m_t$ and true election outcome $\eta_t$. We measure how closely the $m_t$ tracks $\eta_t$ using two complementary metrics: (1) the mean squared error (MSE)  capturing the magnitude of the deviation, and (2) the lag ($\ell$) for which the one-variable regression of the true outcome on the lagged market price has the lowest p-value, indicating if there is a lag introduced between market price and true outcome. Respectively, these measure the presence of a non-random relationship between the true and market values and any lags in this relationship, providing insight into the sensitivity and speed of adjustment of the prediction market. 
\vspace{-1mm}
\subsubsection{Method}

Firstly, we test the performance of our betting market across a range of values for each betting agent attribute. For each agent attribute (stubbornness $s_i$, expertise $e_i$, bias $b_i$, risk aversion $r_i$), we iterate across the parameter range in intervals of 0.1, performing 30 simulations for each parameter value while holding all others constant. For each value of the target parameter, every bettor in our prediction market is initialized with the same constant values shown in \autoref{tbl:bettor_attributes} ($B_{i,0}\sim U(100,1000)$, $V_{i,0}\sim N(0.5,0.05)$, $C_{i,0}=0$, $s_i\sim \mathcal{N}(0.3,0.05)$, $e_i \sim \mathcal{N}(0.9,0.04)$, $b_i \sim 0$, $r_i \sim U(0,1)$). 
Note that, in this set of experiments, we do not include herding dynamics; all agents update independently based solely on their internal valuation process.

When we are testing the effect of varying the spread of budgets $B_{i,0}$ across agents we draw the budget for each agent from a normal distribution with variance $\sigma^2_{B_0}$. This variance is incremented in steps of size \$50 to test the impact of a larger spread in the initial budget of the betting agents.  
\vspace{-1mm}
\subsubsection{Results}\label{sec:abm_valid_result}

\autoref{fig:F1} (B) demonstrates that the accuracy of the betting market, measured as the MSE between the market price and true election outcome, is robust across majority of the parameter value ranges. 

The prediction market is least sensitive to variations in mean levels of risk aversion and budget allocation, with the MSE remaining almost constant in level across all possible values. In the case of both stubbornness and expertise, the MSE only deviates from a stable point at extreme values of each parameter mean. As the value of stubbornness approaches 1, agents approach complete ignorance to any indication of $\eta_t$ causing the market value to deviate significantly from $\eta_t$. Remarkably, the MSE remains close to zero until very high mean values of stubbornness (greater than 0.8). In the case of expertise, the MSE remains nearly constant around 0.05 until the mean value of expertise approaches 1, at which point bettors are receiving near perfect signals of the true election outcome $\eta_t$ such that the MSE of the market price approaches 0. 

However, varying the mean value of bias across the betting population does disturb the predictability of the betting market. As the systematic bias in the beliefs held by the bettors increases or decreases, the MSE of market price increases in magnitude. This result is not surprising as the bias parameter is varied such that it is asymmetric around the true election outcome (i.e. all bettors hold bias in the same direction, either all depreciating or inflating their perception of $\eta_t$). Further research could investigate the effect of a less asymmetric or heterogeneous bias parameter in the betting population. 

These results demonstrate that the information aggregating property of the prediction market ABM exhibits robustness to variation in agent attributes. This has two important implications. First, this demonstrated simulation fidelity suggests that the model of betting agent behavior is sufficiently complex and not sensitive to parameter choices. The model's stability across parameter ranges allows for considerable modeling flexibility providing a strong foundation for future work. Second, the results are consistent with the characterization of prediction markets as systems that aggregate diverse opinions, a core design motivation emphasized in the prediction markets literature \cite{wolfers_prediction_2004}. 

While we cannot validate the ``accuracy'' of prediction markets against realized outcomes until those outcomes occur, we can assess market error ranges relative to observed deviations between leading forecast models and prediction markets. \autoref{fig:F1} demonstrates that the MSE of our market price and $\eta_t$ ranges from 0.006-0.1 which translates to a mean error of \$0.07-\$0.30 between our prediction market and $\eta_t$. For comparison, we use state-level election forecasts from the Economist's election model for the 2016 and 2020 elections, using forecasts published between May 10, 2016 and November 8, 2016 and March 1, 2020 and November 3, 2020, respectively \cite{economist_forecast_data_2020}. The methodology underlying the Economist forecast model is described in \cite{heidemanns2020updated}. Our comparison shows that Polymarket prediction prices deviated from these forecasts by a nearly identical range. We provide visual documentation of this validation benchmark in \autoref{app:market_price_data}.
\vspace{-1mm}
\subsection{Relevance of the Price Error to Electoral Outcomes}

In empirical research, prices in prediction markets are commonly interpreted as the probability of a future event \cite{restocchiOpinionDynamicsExplain2023,wolfers_prediction_2004}. In public settings however, such prices may instead be interpreted using binary decision rules regarding the likely winner, particularly under rapid, headline-level media reporting. Under this interpretation, small deviations around the 0.5 threshold are more likely to flip the implied winner, even when the underlying probabilistic error is small. By treating the market price and true election outcome as normally distributed random variables, we can visualize the probability of misclassification across a range of market and election values \autoref{fig:F1} (D).

\section{An exploration of market manipulation and volatility}\label{sec:application}
The model provides a tool for researchers to evaluate various mechanisms within prediction markets. We provide the first of such an exploration by stress-testing market resilience to price manipulation. In the remainder of this work we explore how high-budget bettors create sustained market price error, and the interaction between such bettors and herding agents.

\subsection{Simulation Study}

\subsubsection{Method}

In order to test the potential for whales to influence the betting market, we simulate the prediction market with 100 betting agents with high expertise (0.95) and introduce a single whale initialized with some proportion of the total budget $\rho_w$. For a given value of $\rho_w$, the whale has around $100 \;\rho_w/(1-\rho_w)$ times the budget of the average non-whale betting agent. The whale has a fixed market valuation above the initial true election outcome. We vary parameter $\rho_w$ in 0.1-size increments to test what, if any, proportion of the total budget ($\times 100$) the whale would require to distort the market price. For each value of $\rho_w$, 100 simulations were performed. As in the experimental results presented in the previous section, we assess the accuracy of the prediction market using MSE and the lag corresponding to the most significant one-variable regression. 

\begin{figure}[ht]
    \centering
    \includegraphics[width=\linewidth]{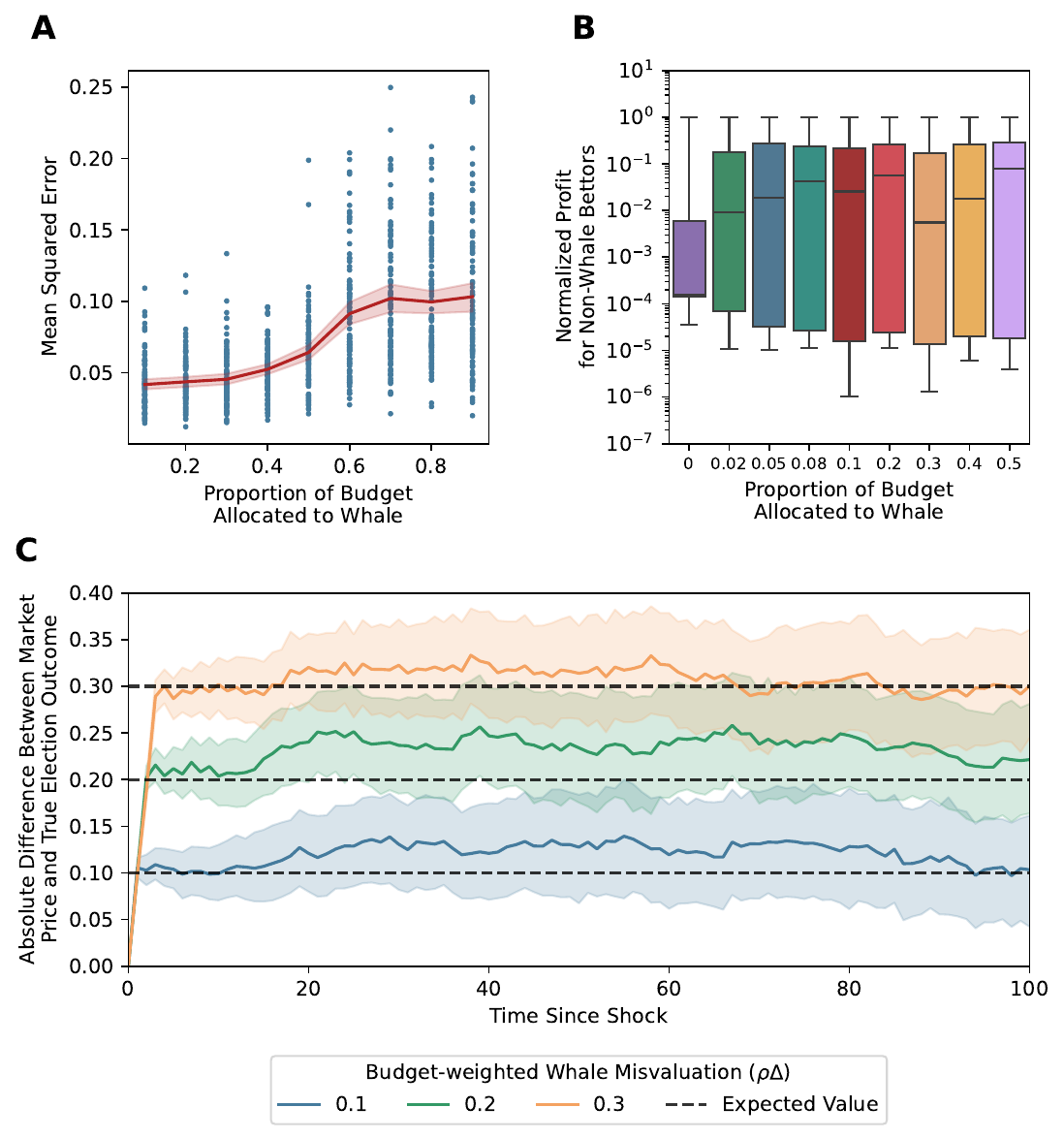}
    \caption{Panel A shows the mean squared error between market price and the proportion of budget allocated to a single whale bettor. The whale has a valuation of 0.6 (error of 0.1 above expected election outcome). As with Figure \autoref{fig:F1} (B), each iteration has 100 agents across 100 timesteps. Here, 100 simulations are performed for each parameter value. All agents are initialized with an expertise of 0.95 to reduce the variance of market price. Default values for other attributes are given in \autoref{tbl:bettor_attributes}. As the proportion of budget allocated to the whale increases, the error increases. Panel (B) shows the profit for non-whale bettors as a proportion of market capital. This plot shows that when a whale is present, the median return for agents with sufficiently high $e_i$ will increase. Panel (C) shows agreement between theoretical error introduced by a large whale ($\rho=0.5$) with varying bias and simulated results from the ABM.}
    \label{fig:whales}
\end{figure}
\vspace{-1.5mm}
Alongside the error introduced by a single whale, we consider how the presence of a whale affects the non-whale agents by considering the median return for non-whale agents adjusted for market capital.

We also consider how stubbornness and herding affect the recovery of the market price after a whale-induced shock. Herding behavior occurs when agents update their internal valuations according to \autoref{eq:valuation_update_herding}. In this setting we consider the absolute error between market price and true election outcome for a system with one whale and 100 non-whale agents. The non-whale agents each have high expertise ($e_i=0.9$) and homogeneous herding weights from the parameter set $h_i=0,\;0.25,\;0.5,\;0.75,\;1$. Whale agents have a fixed valuation set above the market price.
\vspace{-2mm}
\subsubsection{Results}

\autoref{fig:whales} demonstrates the results of introducing a whale into the betting market. We summarize these results below. 

\paragraph{Large Whales Temporarily Distort Market Price} We demonstrate that whales can introduce significant error into the market price, with their effectiveness increasing with budget proportion. Under the outlined parameter set, whales require about 40\% of total market capital to induce meaningful error into the market prices. While the minimum budget threshold for inducing error will vary across parameter sets, our results illustrate that prediction markets exhibit meaningful resilience to manipulation by biased agents but are likely to become vulnerable when such agents control sufficient capital.

We see a similar result whether the whale attempts to manipulate the market price downward or upward. Despite the demonstrated resilience of our underlying prediction market model to changes in bettor behavior, these results demonstrate the potential for whales to nonetheless influence prediction market outcomes. However, our results suggest that the relative size of the whale would need to be extremely large to disturb the prediction market outcome. 

\paragraph{Price Distortions Provide Opportunity for Profit Gain} Additionally, \autoref{fig:whales} (B) demonstrates that a temporary price distortion provides an opportunity for considerable profit gain by well-informed bettors (bettors with high expertise $e_i=0.95$). This profit gain appears to be consistent irrespective of the proportion of total market capital allocated to the whale. This indicates that though the potential distortionary effects of the whale might harm the accuracy of the prediction market, it provides clear opportunity for well-informed bettors to gain profit from the temporarily misaligned market price.

\begin{figure}[H]
    \centering
    \includegraphics[width=0.5\textwidth]{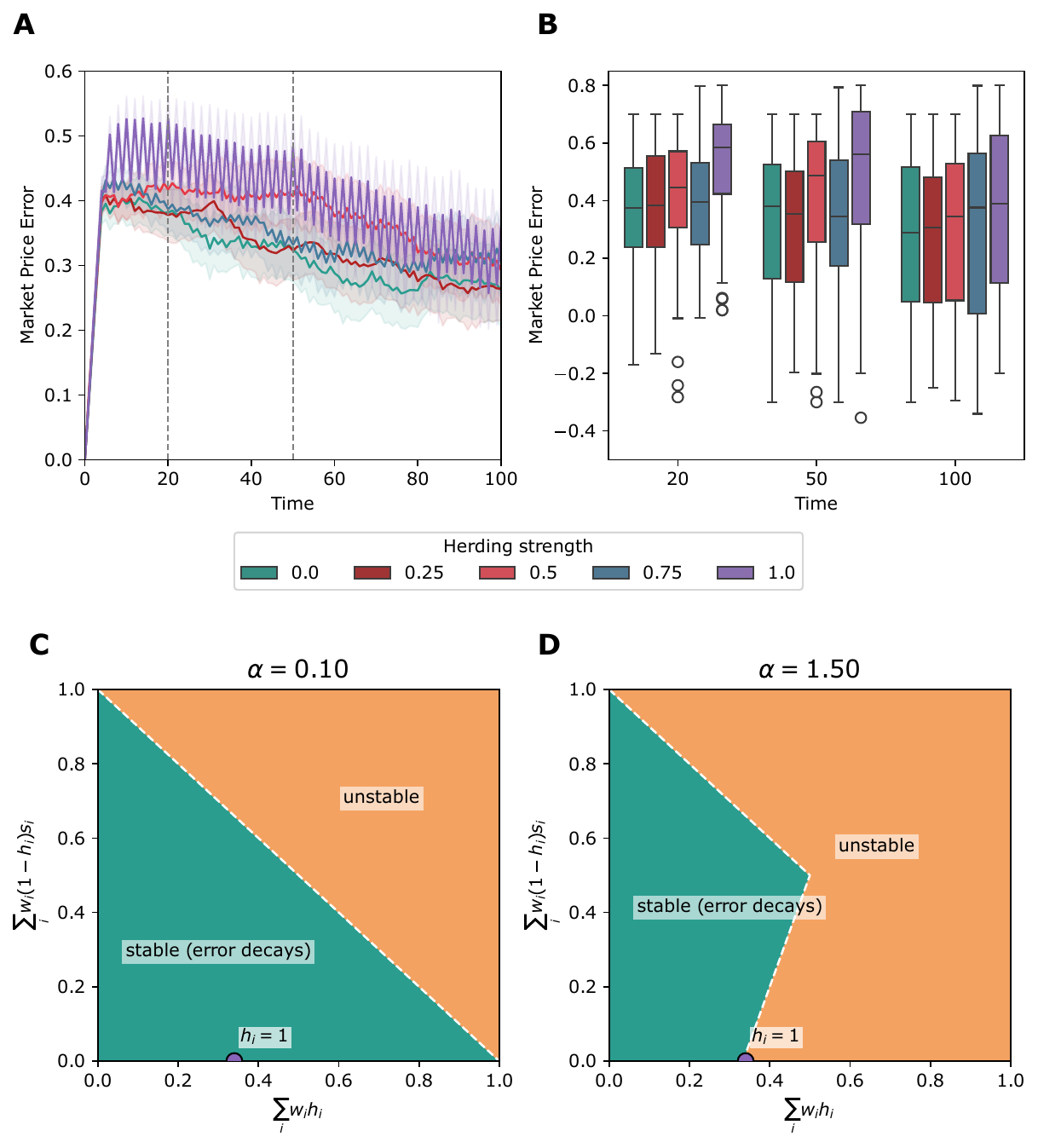}
    \caption{Panel A and B show results for how herding agents impact recovery after a market shock. The system is initialized with a single whale ($\rho=0.3$) and 100 agents with high expertise ($e_i=0.9$ and herding strengths $0,\;0.25,\;0.5,\; 0.75$ and $1$. Instability is visible for $h_i=1$ alongside a decreased rate of decay. Panel A shows the absolute market price error across time for each value of $h_i$ and Panel B shows snapshots at times 20, 50 and 100. For each value of $h_i$, 100 simulations were run. Panel C shows a visualization of phase diagram for some values of $\alpha$. A purple point labeled $h_i=1$ corresponds to the ABM setup with $h_i=1$ in panels A and B.}
    \label{fig:stability_region}
\end{figure}

\paragraph{Herding and High Stubbornness Compound Distortion Magnitude and Duration}
The presence of herding agents influences the decay of a price distortion introduced by a whale. \autoref{fig:stability_region} shows how the price distortion introduced by a whale with 30\% of market capital in a simulated market with 100 agents with homogeneous herding propensities changes as $h_i$ increases. Simulated results in panels (A) and (B) show that small to moderate herding strengths do not greatly affect the rate of market recovery, but large values of $h_i$ can introduce instability characterized by oscillations into the market price. These findings are also explored through theoretical mechanisms in \autoref{sec:theoretical_derivation}.

\subsection{Theoretical Exploration}\label{sec:theoretical_derivation}

To test the results emerging from our simulation-based study, we derive a theoretical solution considering the deterministic behavior of the interacting equations governing our market, its participants, and a high-budget bettor. More precisely, we consider prediction markets as a discrete system where bettor beliefs inform valuations which are transmitted to the market price. By considering the market price update function in \autoref{eq:price_update}, we will consider how internal valuation and herding affect the magnitude of and decay of error between the market price and true election outcome. Using these two settings, we will consider the impact of the presence of a whale-a biased agent with a fixed market valuation and large budget, and the presence of herding agents in the system.

\subsubsection{Absent Herding Behavior}

Consider a setting where agents $i=1,\dots,N$ are \emph{unbiased} with $b_i=0$ and budgets $B_{i,t}=(1-\rho)B_\Omega/N$. To represent a ``whale'', let agent $N+1$ be a \emph{biased} agent with fixed valuation $V_{N+1,t}\equiv W$ and budget $\rho B_\Omega$. Assume holdings $C_{i,t}=0$ for all agents.

Let time $S$ correspond to the system being in the steady state, where \(m_{S+1}=m_S\). When the market price is stable, the net-demand will be zero, and
$$
\sum_{i=1}^{N} x_{i,S}^\star + x_{N+1,S}^\star \;=\; 0.
$$

To determine the impact of a biased agent on this price update function, we use a linearized version of the order size for each agent (\autoref{eq:lin_order_update}) (Further details in \autoref{app:linearised_orders}).

\begin{equation}\label{eq:lin_order_update}
    x_{i,t}^\star \;=\; \frac{B_{i,t}}{m_t(1-m_t)}(V_{i,t}-m_t).
\end{equation}

Now we can use this to explore the bias introduced into the system of agents with $N$ unbiased and 1 biased agent. Assuming the unbiased agents' mean market valuation equals the true value in expectation (i.e. $\mathbb{E}[V_{i,S}]=\eta_S$ under the stated independence assumption and in the absence of bias), we can make the simplification that $V_{i,S}\approx\eta_S$ for each unbiased agent. This gives us,
$$
(1-\rho)B_\Omega(\eta_S-m_S) \;+\; \rho B_\Omega (W-m_S) \;=\; 0.
$$
Assuming that $\delta_S := m_S-\eta_S$ and $\Delta_S := W-\eta_S$,
\begin{equation}\label{eqn:steadystateerror}
   \delta_S \;=\; \rho\,\Delta_S.
\end{equation}

Thus, the steady-state price error equals the biased agent's budget fraction times its valuation error. This gives us the intuitive result that the error introduced into the market price by the presence of a whale is equal to the proportion of the budget which the whale can access and the misvaluation of the whale as validated through our ABM (\autoref{fig:whales} (iii)).

\subsubsection{Allowing for Herding Behavior}
This potential mechanism for market price manipulation is compounded when agents are endowed with herding behavior. We impose herding behavior as described in \autoref{eq:valuation_update_herding}.

When the market price and true election outcome agree, this herding behavior ($h_i>0$) corresponds to increased agreement between agent valuation and market price, since if $m_t\approx \eta_t$ and conditional on $V_{i,t}$ and $\eta_t$, the distribution of $V_{i,t+1}\mid V_{i,t},\eta_t$ is normal with mean $$(1-h_i)\big((1-s_i)(\eta_t-b_i)+s_iV_{i,t}\big) + h_i \eta_t,$$ and variance $$(1-h_i)^2(1-s_i)^2(1-e_i).$$

For $0 < h_i\leq 1$, the variance of this distribution is reduced compared to $h_i=0$.

Now consider the system where a shock introduces error between the market price and the true election outcome (i.e. $m_t=\eta_t + \delta_t$). The shock may occur in the presence of a whale, and we consider how herding agents change how quickly error is removed from the system. In a system of $N$ agents, the budget-weighted valuation $\bar{V}_t$ has an expected error of
$$
\mathbb{E}[\bar{V}_{t+1}-\eta_{t+1} \mid V_{i,t}, \eta_t] = \sum_{i=1}^N w_i \big[(1-h_i) s_i ( V_{i,t}-\eta_t) + h_i \delta_t \big]
$$
where $w_i = \dfrac{B_i}{\sum_{i=1}^N B_i}$. In this budget-weighted average, the average error when $\delta_t$ is small shrinks by $\sum_{i=1}^N w_i (1-h_i) s_i$. So we can see that when the error between the true election outcome and market price is small and agents exhibit more herding and less stubbornness, any error between the internal valuation and true election outcome drop quickly.

For very small $\delta_t$, the expected deviation between an individual agent's internal valuation and the truth is shrinks with factor $(1-h_i)s_i$. For small values of $s_i$ or large values of $h_i$, the internal valuation of the agents corrects to the true value more quickly.  A more complete exploration is given in the Supplementary Material (\autoref{app:linearised_orders}).

In general, we can see that large herding weights $h_i$ increase the strength of the feedback from the market price back into valuations. Large values of $s_i$ cause the market error to correct more slowly. The interplay of these values with the market step size $\lambda$ are important, and may result in unstable or oscillatory behavior if the internal valuation of the agents tends to over correct.

\section{Discussion}

In this work, we build an open-source agent-based model of a prediction market which we use to explore the potential for market manipulation by betting agents with considerable market capital. We find that ``whales'' are able to temporarily distort price levels in the market. Centrally, we determine that the degree and duration of this distortion depends on the relative budget of the whale agent, stubbornness of non-whale bettors, and the degree to which individuals engage in herding behavior. A theoretical analysis, given in \autoref{sec:theoretical_derivation} supports these findings and identifies explicit relationships between agent attributes and market error and duration.

Recent examples of individuals placing outsized bets in prediction markets have raised questions around the potential consequences for real-world elections. Though our model indicates the potential for diverse agent behavior and competition to self-regulate the market price, we identify conditions under which such manipulation could interfere with these resilient dynamics. Even a temporary distortion caused by a large, stubborn, or strategically biased whale could shape voter expectations, campaign donations, or media narratives. In such cases, market manipulation would not merely affect market accuracy but could feed back into the political process itself, influencing the very outcome being predicted. While we do not incorporate non-market incentives into our whale agent specification, contexts like the U.S., where there is an established pattern of wealthy actors deploying substantial campaign finance contributions to influence elections suggest that such motivations are logically plausible.

Market design and enforcement constrain or enable conditions under which whale-driven price distortions can arise and persist. Earlier prediction markets, such as the Iowa Electronic Markets and PredictIt, typically imposed per-trader caps in the range of hundreds to a few thousand USD per market, which prevented whales through market design. The U.S.-regulated Kalshi platform does not impose a single global position limit; instead, position limits are specified at the contract level, with some markets permitting multi-million-dollar positions for eligible participants under identity-verified participation and CFTC oversight. The current large Polymarket platform represents a further shift in market conditions: participation is pseudonymous and cross-jurisdictional through cryptocurrency use, with no binding per-trader caps, while its price signals are globally visible in real time. Where legal enforcement is limited or fragmented, mitigating manipulation relies more heavily on market design choices. 

Our results point to the need for a thoughtfully considered regulatory approach, mainly concerning maximum order sizes, to avoid market price manipulation in election contexts where legal enforcement is feasible.  In the US, the Commodity Futures Trading Commission (CFTC) currently oversees federal regulation of prediction markets and has taken a deregulatory approach to election betting in recent years. In a 2024 decision, a federal judge ruled in favor of plaintiff Kalshi, permitting election betting a month prior to the federal election \citep{united_states_court_of_appeals_for_the_district_of_columbia_circuit_kalshiex_2024}. Within days, Kalshi advertised the legality of election betting on their platform and touted their capacity to handle ``individual trades of up to \$7 million per contract, while eligible contract participants (ECPs), such as corporations and investment firms can trade up to \$100 million'' \citep{mansour_its_2024}. Recently, the CFTC has moved to drop its appeal in this case, further undermining efforts to regulate prediction markets \citep{matthews_cftc_2025, thomson_reuters_practical_law_cftc_2025}. Contrary to this trend, our findings suggest the need for a regulatory framework that strikes a balance between recreational freedom and safeguarding democratic integrity.

\subsection{Limitations and Further Work}

\subsubsection{Public Opinion and Belief Formation}The betting agents in our model update their market valuations based on a ``true election value.'' While unrealistic, this mechanism allows for a simple implementation of a real-world process principally characterized by imperfect or ``fuzzy'' information. Future work could incorporate more realistic information environments wherein agents integrate global, local, and/or social information. For example, embedding a social network of voting agents within the model, betting agents could formulate beliefs based on their visibility of various communities across the network. Such extensions would not only approach greater realism, but also enable investigation of how misinformation, unbalanced media amplification, and polarization affect information flows from voting constituency to market. This could also enable the use of prediction markets as potential diagnostic tools for understanding evolving social dynamics in electoral contexts. 

\subsubsection{Market-Election Feedback} Further work should explore the potential feedback between prediction markets and voting behavior. Any non-zero likelihood of market manipulation could have significant implications for the integrity of democratic elections or other real-world outcomes upon which betting markets speculate. Should there exist a potential channel through which a prediction market price could influence voting behavior, public endorsements, campaign finance contributions, or other levers through which a voting population, a political establishment, or a private interest group might affect the outcome of an election, it would raise important questions about the interplay between market-based forecasts and democratic processes. This work does not explore such an interaction. Whether such a link exists would determine the degree of policy concern warranted in light of the results presented in this work. 

\subsubsection{Bettor Intentions and Non-Pecuniary Utility} In this work, we utilize a complex systems approach, employing a simulation based model to isolate relationships between bettor attributes, behavior and market distortions. This approach does not consider agent intention. Future work could investigate the degree to which agents experiencing non-market utility gains from a political outcome could influence the incentives of biased agents. Careful crafting of such a mechanism could allow whale-like behavior to emerge endogenously. The prevalence of substantial campaign finance contributions deployed to influence electoral results across federal and state jurisdictions in the United States suggests that such incentives may operate within online prediction markets as well.

\subsubsection{Market Design} Our ABM implements a random-order, price-first matching prediction market. Alternative market designs include automated market makers or call-market clearing mechanisms. While market design and other modeling choices may affect quantitative outcomes, our implementation reflects the dominant microstructure used in the contemporary prediction market literature and in practice, and results should be interpreted in that context.

\section{Acknowledgments}
We would like to acknowledge thoughtful comments and discussions with fellow participants and faculty mentors of the 2024 Complexity Global School in Bogotá, Colombia. In particular, we would like to thank faculty mentor Rajiv Sethi for early motivating discussions. We would also like to thank Travis Holmes, Jordan Kemp, Will Tracy, and Renaud Lambiotte for important input in later stages of the project. 

\section{Funding}
This research was supported by the Emergent Political Economies grant through the Omidyar Network. We would like to thank the Santa Fe Institute for allowing us to conduct this research in the context of the 2024 Complexity Global School in Bogotá, Colombia and 2025 Postdocs in Complexity Conference at the Santa Fe Institute in Santa Fe, New Mexico, USA.

\newpage
\section{Author contributions statement}
\textbf{Conceptualization}: B.S., E.M., A.B., J.W.; \textbf{Methodology}: B.S., E.M., A.B., J.W.; \textbf{Software}: B.S., E.M.; \textbf{Validation}: B.S., E.M.; \textbf{Formal analysis}: B.S.; \textbf{Investigation}: B.S., E.M., A.B., J.W.; \textbf{Resources}: N/A; \textbf{Data Curation}: B.S., E.M.; \textbf{Writing - Original Draft}: B.S., E.M., J.W.; \textbf{Writing - Review \& Editing}: B.S., E.M., A.B., J.W.; \textbf{Visualization}: B.S., E.M.; \textbf{Supervision}: N/A; \textbf{Project administration}: B.S., E.M., A.B., J.W.; \textbf{Funding acquisition}: B.S., E.M., A.B., J.W..

\bibliographystyle{plain}
\bibliography{reference}
\newpage
\section{Supplementary Material}

\subsection{Order update function in the steady state}\label{app:linearised_orders}

Let time $S$ correspond to the system being in the steady state, where \(m_{S+1}=m_S\). When the market price is stable, the net-demand will be zero, so
$$
\sum_{i=1}^{N} x_{i,S}^\star + x_{N+1,S}^\star \;=\; 0.
$$

To determine the impact of a biased agent on this price update function, we start by finding the expected order size for an agent. Possible orders correspond to maxima of \autoref{eq:EU_fun}, so begin by taking the derivative with respect to $x$ (\autoref{eq:crra_utility_fun}),
\begin{align}\nonumber
0 \;=\; &V_{i,t}(1-m_t)\,u'\big(B_{i,t}-m_t x + C_{i,t}+x\big)
\;\\& +\;(1-V_{i,t})m_t\,u'\big(B_{i,t}-m_t x\big).\label{eq:FOC_CRRA}
\end{align}

The first derivative of the utility function is $u'(\omega)=\omega^{-r_i}$ for the non-log case ($r_i\neq 1$), and $u'(\omega)=\omega^{-1}$ for the log case ($r_i= 1$). For general \(r_i\neq1\), \autoref{eq:FOC_CRRA} is a nonlinear equation in $x$,
\begin{align*}
    0&=V_{i,t}(1-m_t)\big(B_{i,t}-m_t x + C_{i,t}+x\big)^{-r_i} \\&+ (1-V_{i,t})m_t\big(B_{i,t}-m_t x\big)^{-r_i}.
\end{align*}

In the log-case the solutions are\begin{equation}\label{eq:log_xstar}
x_{i,t}^{\star} \;=\; \dfrac{m_t\big(B_{i,t}+C_{i,t}-C_{i,t}V_{i,t}\big)\;-\;B_{i,t}V_{i,t}}{m_t( m_t -1)}.%
\end{equation}

To explore the theoretical error introduced by a biased agent, we will use assume that $C_{i,t}=0$. When \(C_{i,t}=0\) the expression reduces to the expression shown in \autoref{eq:lin_order_update}.

Note, we assume \(m_t\notin\{0,1\}\) and that $x_{i,t}^\star$ is a valid solution with respect to the given constraints.

When $C_{i,t}\neq0$ or $r_i\neq 1$, we can use a linearized version,

$$
x_{i,t}^\star(V_{i,t}) \approx x_{i,t}^\star(m_t) + \left.\frac{\partial x_{i,t}^\star}{\partial V}\right|_{V=m_t}(V_{i,t}-m_t).
$$

when $\left.\frac{\partial x_{i,t}^\star}{\partial V}\right|_{V=m_t}$ exists and $|V_{i,t}-m_t|$ is small.

\subsection{Variance of error in the market price}
In the specified ABM, each agent has an expertise which controls how clearly they can access information about the market price. This expertise influences the variance of the market price, impacting the uncertainty of the market price around the true election outcome. 

Consider a setting with $N$ agents, each with bias $b_i=0 $, budget $B_{i,t}$ and internal valuation $V_{i,t}$.  Using the linear order updates for an unbiased agent, we have that 
$$
x_{i,t}^\star \;=\; \frac{B_{i,t}}{m_t(1-m_t)}(V_{i,t}-m_t).
$$

So if we condition on fixed values of $B_{i,t}, m_t$, 
\begin{align*}
x_{i,t}^\star \mid B_{i,t}, m_t, V_{i,t-1}, b_i=0 \sim \mathcal{N} \bigg( \beta, \beta^2 (1-e_i) \bigg)
\end{align*}

for $\beta := \dfrac{B_{i,t} \eta_t}{m_t(1-m_t)}((1-s_i)+s_i V_{i,t-1}-m_t)$.

If we consider the net demand across all agents, conditioning on $\{B_{i,t}\}$, $\{m_t\}$, and $\{V_{i,t-1}\}$, we have that the distribution of $D_t$ is approximately
\begin{align*}
D_t \;\sim\; \mathcal{N}\!\bigg(&
\sum_{i=1}^N \dfrac{B_{i,t} \eta_t}{m_t(1-m_t)}\big((1-s_i)+s_i V_{i,t-1}-m_t\big),\\ &
\sum_{i=1}^N \left(\dfrac{B_{i,t}}{m_t(1-m_t)}\right)^{\!2}(1-s_i)^2(1-e_i)
\bigg),
\end{align*}

where we have used the independence of agents’ signals.

From this, the variance of the net demand is
\begin{equation}\label{eq:var_D}
\mathrm{Var}(D_t) = \sum_{i=1}^{N} 
\left(\dfrac{B_{i,t}}{m_t(1-m_t)}\right)^{\!2}
(1-s_i)^2(1-e_i).
\end{equation}

If the price is a random walk with variance $\sigma^2_\eta$, then the variance of the error between the market price and the true election outcome is a function of the variance of the net demand. 

\autoref{eq:var_D} and \autoref{eqn:steadystateerror} demonstrate that the budget-weighed bias of the agents can introduce bias into the market price, while expertise controls the variance of the error between the market price and true election outcome.

\subsection{Stability of price error under herding conditions}\label{app:herding_stability}

To demonstrate the effect which herding has on the market price error over time, assume the linear order update as given in \autoref{eq:lin_order_update} to find a linear approximation of the error between market price and true election outcome,
\begin{align*}
    \delta_{t+1} = m_{t+1} - \eta_{t+1} &= m_t + \lambda \dfrac{D_t}{K_t} - \eta_{t+1} \\ 
    &\approx  m_t + \alpha (\bar{V_t}-m_t) -\eta_t \\
    &= (1-\alpha) \delta_t + \alpha ( \bar{V}_t - \eta_t).
\end{align*}

Here, $\alpha \propto \lambda m_t(1-m_t) \leq 0.25 \lambda$. This assumption holds when $r_i=1$ and the market price is sufficiently close to the agent's valuation.

Combining this with the expected budget weighted valuation across the agents, we can express the market error as an AR(2), with 
\begin{align*}
\delta_{t} = &\big[(1-\alpha)+\sum_i w_i (1-h_i)s_i \big]\,\delta_{t-1} \\&+ \big[\alpha \sum_i w_i h_i - (1-\alpha) \sum_i w_i (1-h_i)s_i \big]\,\delta_{t-2}.
\end{align*}

Analysis of this AR(2) model gives more precise conditions for when the error between true election outcome and market price will decay to zero. However, in general the error will decay if $1 > \sum_{i=1}^N w_i [h_i + (1-h_i) s_i]$. For an agent population, this sum can at most be equal to 1, for example if $h_i=1, s_i=0$ for all agents. Hence, for all reasonable parameter values, we will expect the error between the market price and true election outcome to decrease. See \autoref{fig:stability_region} for a more precise visualization of the regions in which the error will decay to zero. A more detailed analysis follows.

The coefficient on $\delta_{t-1}$$, $$(1-\alpha)+\sum_i w_i(1-h_i)s_i$ is non-negative because $\alpha, w_i, h_i, s_i \geq 0$. Note however that being non-negative does not guarantee stable or fast decay. If this coefficient exceeds 1 or when the two coefficients violate stationarity conditions the system can be oscillatory or slowly decaying.

The sign of the coefficient on $\delta_{t-2}$,
  $$
\alpha\sum_i w_i h_i - (1-\alpha)\sum_i w_i(1-h_i)s_i,
$$ can determine if the error dampens or oscilates in time. If the coefficient is smaller than 0 (equivalently
$\alpha\sum_i w_i h_i < (1-\alpha)\sum_i w_i(1-h_i)s_i$,
then $\delta_{t-2}$ term tends to dampen the error. In our system, $\alpha$ tends to be small, so this is generally true.

If the coefficient is greater than zero, for example for some combinations of large $\alpha$, large $h_i$ or small $s_i$, the past error tends to reinforce the newer error, increasing persistence and the risk of oscillation or growth. By comparing the coefficient in the AR(2) system, we can see that when feedback from market price into valuations is weak (large  $h_i$ or $s_i$ small) relative to valuation carry-over then the error induced by the shock decays smoothly. Otherwise, errors will tend to persist or oscillate.

\subsubsection{Stability analysis}
To derive the stability of AR(2) model for the error between the market price and true election outcome, let $$S_w=\sum_{i=1}^N w_i(1-h_i)s_i,$$ and the budget-weighted herding strength average be $$\bar{H_B}=\sum_{i=1}^N w_i h_i$$. Set
$$
a := (1-\alpha) + S_w,\qquad
b := \alpha \bar{H_B} - (1-\alpha)S_w.
$$
The error between the market price and true election outcome is then,
$$
\delta_t = a\,\delta_{t-1} + b\,\delta_{t-2},
$$
with characteristic polynomial
$$
r^2 - a r - b = 0.
$$

The two roots are
\begin{align*}
    r_{1,2} \;&=\; \frac{a \pm \sqrt{a^2 + 4b}}{2}
    \\&=\;
\frac{(1-\alpha)+S_w \pm \sqrt{\big((1-\alpha)-S_w\big)^2 + 4\alpha \bar{H_B}}}{2}.
\end{align*}

For a second-order discrete-time polynomial (characteristic polynomial)
\[
r^2 - p_1 r - p_2 = 0
\]
(with \(p_1=a,\ p_2=b\)), the necessary and sufficient conditions for the solution to be stable are,
$$
\begin{aligned}
1 + p_1  & > p_2,\\
1 - p_1 &> p_2,\\
p_2 &> -1.
\end{aligned}
$$

The conditions for stability are,

$$
\begin{aligned}
\text{(i)}\quad & 1 - a  > b \quad\Longleftrightarrow\quad 1 > \bar{H}_B +S_w
 \\[6pt] 
\text{(ii)}\quad &1 + a > b 
\quad\Longleftrightarrow\quad 2> \alpha + \alpha \bar{H}_B + (\alpha-2) S_w 
,\\[6pt] 
\text{(iii)}\quad &b \;>\; -1
\quad\Longleftrightarrow\quad\alpha > \dfrac{S_w-1}{\bar{H}_B + S_w}. 
\end{aligned}
$$

We know that $0\leq \alpha \propto \lambda\; m_t(1-m_t) \leq 0.5 \lambda$, so we expect $\alpha$ to be small.

Since $S_w=\sum_{i=1}^N w_i(1-h_i)s_i$, and the budget-weighted herding strength average be $\bar{H_B}=\sum_{i=1}^N w_i h_i$, for $w_i, s_i, h_i \in [0,1]$, $S_w, \bar{H_B} \in [0,1]$.

From these constraints, we see:

\begin{enumerate}
    \item Since \(\alpha>0\) in the model, (i) implies $1 > H^B + S_w$. 
    \item For $\alpha < 1$, constraint (ii) will always be satisfied. 
    \item $S_w-1\leq 0$ so (iii) is generally trivial, but care should be taken if $S_w$ and $\bar{H}_B$ are both extremely small or zero.
\end{enumerate}

In simulations, generally $\alpha < 1$, since $\lambda$ is small. This means the most informative condition is (i). 

\subsection{Validation of Market Price Deviation}\label{app:market_price_data}

The following displays the RMSE between the time series of PredictIt market prices (daily closing share price) and Economist forecasts (weekly forecast) of state-level presidential election returns in 2016 (May 10 - November 8) and 2020 (March 1-November 3). We reconcile the periodicity discrepancy between the two data sources by comparing the closing PredictIt share price to the weekly forecast published on the same date. The range of differences between the prediction market price and Economist forecasts are between 5-30 percentage points, indicating stark variability in predictive accuracy. However, this range provides an important benchmark for assessing a feasible price deviation range between our simulated prediction market price and the true electoral preferences.

\begin{figure*}
    \centering
    \includegraphics[width=0.8\textwidth]{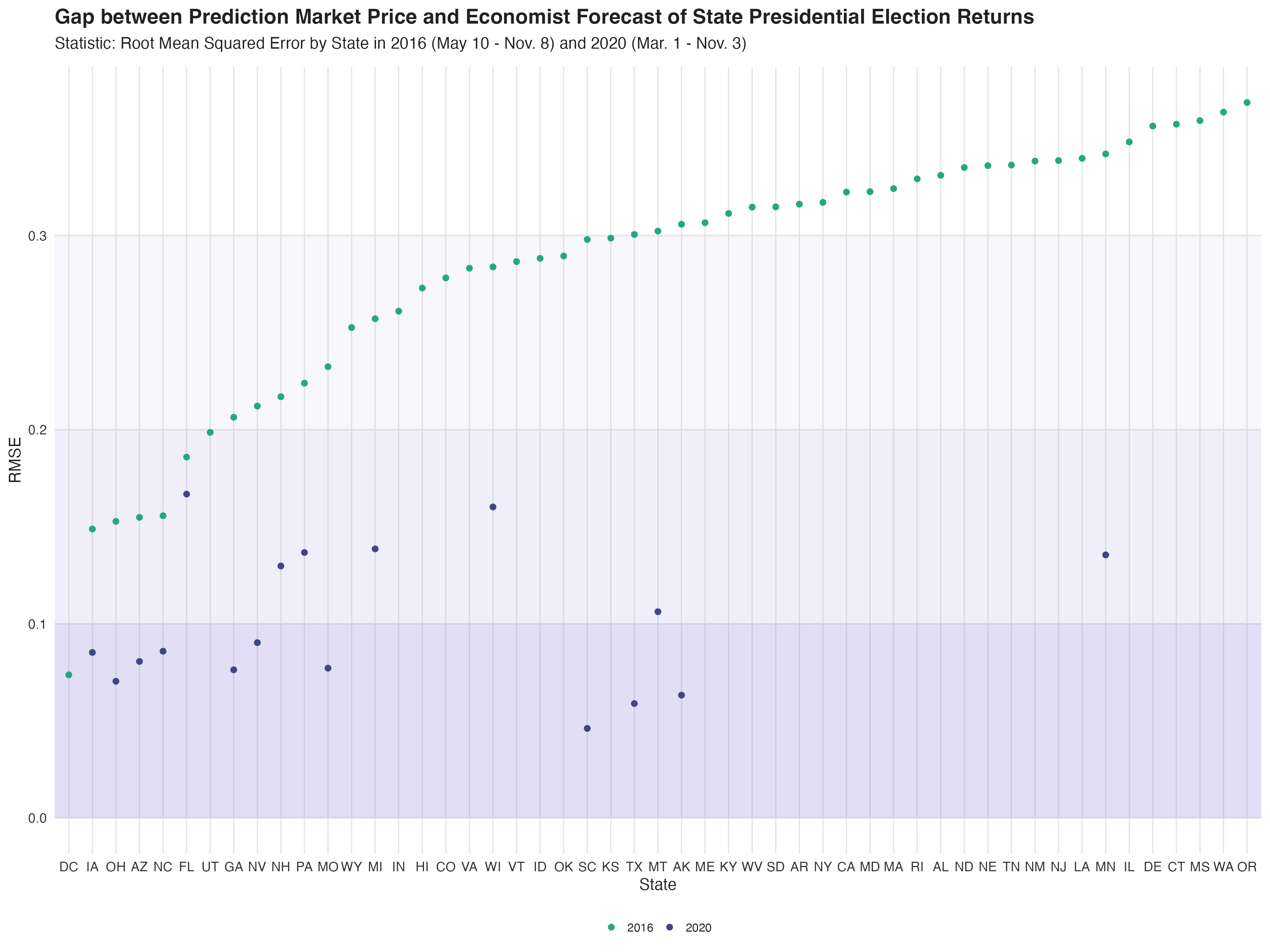}
    \caption{Root mean squared error between Polymarket market price and Economist Forecast Model of state election returns in the 2016 and 2020 presidential elections.}
    \label{fig:market_price_comparison_data}
\end{figure*}

\begin{figure*}
    \centering
    \includegraphics[width=0.8\textwidth]{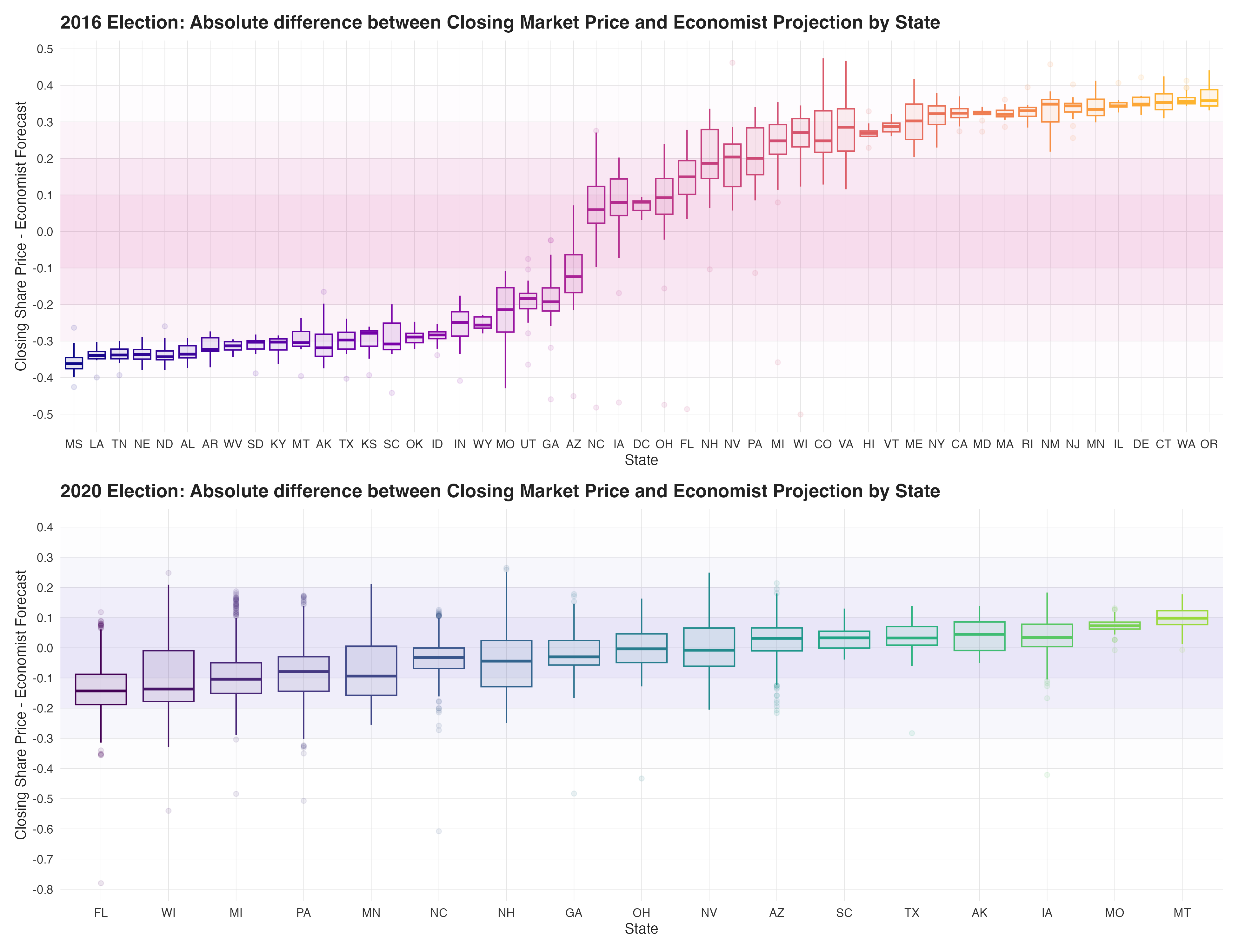}
    \caption{Absolute difference between Polymarket market price and Economist Forecast Model of state election returns in the 2016 and 2020 presidential elections.}
    \label{fig:market_price_comparison_data_absolute}
\end{figure*}

\begin{figure*}
    \centering\includegraphics[width=0.8\textwidth]{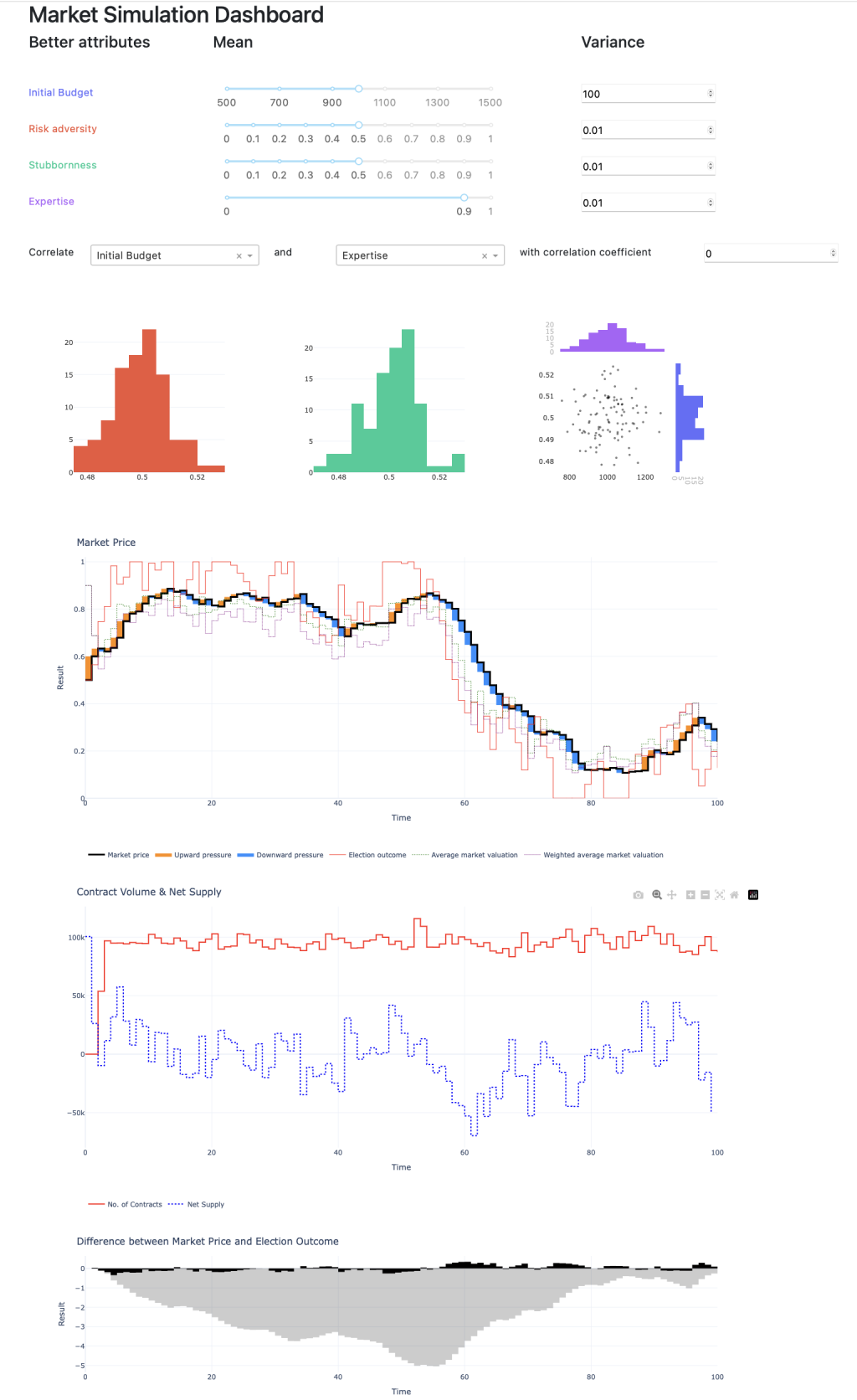}
    \caption{Example output from the Dash Application. In this example, each bettor attribute is drawn from a normal distribution with user-specified values. The user can correlate samples for two of these variables, visualizing how these changes affect the prediction market.}
    \label{fig:dashapp}
\end{figure*}

\end{document}